\documentclass{aa}

\usepackage{graphicx}
\def\chandra{{\it Chandra\,}}
\def\xmm{{XMM-{\it Newton\/}}}
\newcommand{\cgs}{${\rm erg~cm}^{-2}~{\rm s}^{-1}$}
\newcommand{\lum}{\rm erg~s$^{-1}$}
\def\lesssim{\,\lower2truept\hbox{${<\atop\hbox{\raise4truept\hbox{$\sim$}}}$}\,}
\def\gtrsim{\,\lower2truept\hbox{${>\atop\hbox{\raise4truept\hbox{$\sim$}}}$}\,}

\begin{document}

   \title{\chandra and ALMA observations of the nuclear activity in two strongly lensed star forming galaxies}
   \authorrunning{M. Massardi et al.}
   \titlerunning{\chandra and ALMA observations of SDP.9 and SDP.11}
   \subtitle{}

\author{M. Massardi\inst{\ref{inst1}}\and A. F. M. Enia\inst{\ref{inst2}}\and M. Negrello\inst{\ref{inst3}}\and 
 C. Mancuso\inst{\ref{inst1}\ref{inst4}\ref{inst5}\ref{inst6}}\and A. Lapi\inst{\ref{inst4}\ref{inst5}\ref{inst6}}\and\\  C. Vignali\inst{\ref{inst7}\ref{inst8}}\and R. Gilli\inst{\ref{inst7}}\and S. Burkutean\inst{\ref{inst1}} \and L. Danese\inst{\ref{inst4}\ref{inst5}\ref{inst6}}\and G. De Zotti\inst{\ref{inst9}}}

\institute{INAF, Istituto di Radioastronomia - Italian ARC, Via Piero Gobetti 101, I-40129 Bologna, Italy \email{massardi@ira.inaf.it}\label{inst1}
\and
Dipartimento di Fisica e Astronomia, Università di Padova, vicolo dell'Osservatorio 2, I-35122, Padova, Italy\label{inst2}
\and
School of Physics and Astronomy, Cardiff University, The Parade, Cardiff CF24 3AA, UK\label{inst3}
\and
Astrophysics Sector, SISSA, Via Bonomea 265, I-34136 Trieste, Italy\label{inst4}
\and
INAF-Osservatorio Astronomico di Trieste, via Tiepolo 11, 34131 Trieste, Italy\label{inst5}
\and
INFN-Sezione di Trieste, via Valerio 2, 34127 Trieste, Italy\label{inst6}
\and
Dipartimento di Fisica e Astronomia, Università degli Studi di Bologna, viale Berti Pichat 6/2, 40127 Bologna, Italy\label{inst7}
\and
INAF, Osservatorio Astronomico di Bologna, via Ranzani 1, 40127 Bologna, Italy\label{inst8}
\and
INAF, Osservatorio Astronomico di Padova, Vicolo Osservatorio 5, I-35122 Padova, Italy\label{inst9}
}
   \date{}


  \abstract
   {}
   {According to coevolutionary scenarios, nuclear activity and star formation play relevant roles in the early stages of galaxy formation. We aim at identifying them in high redshift galaxies by exploiting high-resolution and sensitivity X-ray and millimetre-wavelength data to confirm the presence or absence of star formation and nuclear activity and their relative role in shaping their SEDs and contributing to their energy budget.}
   {In the current paper we present the data, model and analysis in the X-ray and mm bands for two strongly lensed galaxies, SDP.9 (HATLAS J090740.0-004200) and SDP.11 (HATLAS J091043.1-000322) that we selected in the Herschel-ATLAS catalogues as having an excess emission in the mid-IR regime at redshift $\gtrsim1.5$, suggesting the presence of a nuclear activity in the early stages of galaxy formation. We observed both of them with \chandra ACIS-S in the X-ray regime and analyzed the high resolution mm data available in the ALMA Science Archive for SDP9, and, by combining the information available in mm, optical and X-ray bands we reconstructed the source morphology.}
   {Both the targets were detected in the X-ray, strongly indicating the presence of highly obscured nuclear activity. 
   
   ALMA observations for SDP9 for continuum and CO(6-5) spectral line with high resolution (0.02arcsec corresponding to $\sim65$ pc at galaxy distance) allowed us to estimate the lensed galaxy redshift to a better accuracy than pre-ALMA estimates (1.5753$\pm0.0003$) and to model the emission of the optical, millimetric, and X-ray band emission for this galaxy. We demonstrated that the X-ray emission is generated in the nuclear environment and it strongly support the presence of nuclear activity in this object. On the basis of the X-ray data, we attempted an estimate of the BH properties in these galaxies.    }
  {By taking advantage of the lensing magnification we identified weak nuclear activity associated with high-$z$ galaxies with large star formation rates, useful to extend the investigation of the relationship between star formation and nuclear activity to two intrinsically less luminous, high-$z$ star forming galaxies than was possible so far. Given our results only for two objects, they solely cannot constrain the evolutionary models, but provide us with interesting hints and set an observational path towards addressing the role of star formation and nuclear activity in forming galaxies.}


   \maketitle
%

\section{Introduction}

Modern scenarios of co-evolution between supermassive black holes (BHs) and their host galaxies envisage star formation and BH accretion to be in situ, time-coordinated processes (e.g., Lapi et al. 2006, 2011, 2014; Lilly et al. 2013; Aversa et al. 2015; Mancuso et al. 2016a,b), triggered by the early collapse of the host dark matter halos, but subsequently controlled by self-regulated baryonic physics and in particular by energy/momentum
feedbacks from supernovae/stellar winds and AGNs. This results in a distinctive evolutionary pattern for the emission associated with the stellar and the nuclear component.

In a nutshell, during the early stages of a massive galaxy evolution, huge gas reservoirs can sustain violent, almost constant SFRs $\dot M_\star\ga 10^2\, M_\odot$ yr$^{−1}$, while the ambient medium is quickly enriched with metals and dust; in such conditions, the galaxy behaves as a bright sub-mm/far-IR source. After a time $\sim \hbox{several}\times10^8$ yr the SFR is abruptly quenched by the energy/momentum feedback from the central supermassive
BH, and the environment is cleaned out; thereafter the stellar populations evolve passively and the galaxy becomes a red and dead (early-type) spheroid.

From the point of view of the central BH, during the early stages plenty of gas is available from the surroundings, so that considerable accretion rates can sustain mildly super-Eddington emission. However, during these early stages the BH bolometric luminosity is substantially smaller than that of the host galaxy, so that the dust-enshrouded AGN may be observable only as a faint X-ray nucleus. However, the nuclear power in self-regulated conditions increases exponentially and after some $10^8$ yr progressively attains or even exceeds that from star formation in the host; interstellar gas and dust are removed and star formation is appreciably reduced or quenched, so that the overall system behaves as an optical quasar. Residual gas present in the central regions of the galaxy can be accreted onto the BH at progressively lower, sub-Eddington accretion rates, as observed on the average in unobscured AGNs.

The coevolution scenario sketched above has been extensively tested (see Lapi et al. 2014, 2017; Mancuso et al. 2016a,b,2017 and references therein) on a wealth of data including : SFR/stellar mass functions of galaxies and AGN luminosity/BH mass functions; IR/sub-mm/radio counts and redshift distributions; main sequences of star-forming galaxies (SFR vs. stellar mass correlation) and AGNs (nuclear X-ray power vs. SFR or stellar mass), redshift evolution of the cosmic SFR and BH mass density. However, the picture requires further validation especially in the early stages of the coevolution, that plainly are not easy to pinpoint given the sensitivity and resolution limits of current X-ray and far-IR facilities.

Fortunately, strong gravitational lensing by foreground objects offers an extraordinary potential to advance our understanding of such elusive early stages. It not only yields flux boosting that can reach factors $\mu\ge 10$, allowing us to explore regions of the luminosity/redshift space that would otherwise be unaccessible, but also stretches the images by factors $\simeq \mu^{1/2}$, allowing the study of fine spatial details.

In fact, observations have shown that almost all high-redshift ($z>1$) dust-obscured star forming galaxies selected in the sub-millimetre (submm galaxies) with flux density above $\sim 100$ mJy at 500 $\mu$m are gravitationally lensed by a foreground galaxy or a group/cluster of galaxies (see Negrello et al. 2010, 2014, 2017; Wardlow et al. 2013; Nayyeri et al. 2016). These submm bright sources are rare ($<0.3$ deg$^{-2}$ at $S_{500\mu m}>100$ mJy, Negrello et al. 2007) and therefore only detectable in wide-area submm surveys, like those by the Herschel Astrophysical Terahertz Large Area Survey (H-ATLAS, Eales et al. 2010). The detected $z>1$ objects are generally dusty galaxies with star formation rates $>10^2\, M_\odot$ yr$^{-1}$, occurring in a strongly dust-enshrouded environment and thus corresponding to FIR luminosity $L_{\rm FIR}>10^{12}\, L_\odot$.

Some of them show excess emission in the mid-IR (Negrello et al. 2014) if compared to the commonly used SEDs of galaxies with intense star formation. This excess may be accounted for by nuclear activity (Fritz et al. 2006), but the mid-IR data alone are not sufficient to draw firm conclusions about the presence of an AGN, let alone its characterization, because they indicate that AGNs, if present, are weak. Furthermore, the SED fitting requires photometric information over a broad spectral range without which the solutions might strongly depend on the used fitting templates.

On the one hand, high-resolution and sensitivity X-ray data might unambiguously confirm the presence or absence of a significant AGN emission. On the other, high-resolution and sensitivity mm-wavelengths data are precious to identify the dust and gas properties associated with both the star formation process and tracing the nuclear feeding and feedback processes dynamics. By combining them we can estimate the physical properties of the galaxy components and the relative roles of star formation and nuclear activity in shaping their SEDs and contributing to their energy budget.

Observations established, beyond any reasonable doubt, that the growths of galaxies and super-massive black holes are intimately connected by investigating the nuclear activity in X-rays in galaxies with large SFRs mainly selected at FIR or submm wavelengths (e.g. Mullaney et al. 2012, Wang et al. 2013, Delvecchio et al. 2015) or by studying the star formation in AGN host galaxies, exploiting the follow-up at FIR and (sub)mm wavelengths of X-ray selected AGNs and of optically selected QSOs (e.g. Mullaney et al. 2012b; Page et al. 2012; Santini et al. 2012)).

With a similar approach we confirm the properties of the nuclear activity in two lensed galaxies, detected in the Herschel-ATLAS catalogues, characterized by large SFRs, by means of X-ray \chandra ACIS-S observations and exploiting high resolution mm data available in the ALMA Science Archive. With these new data we improved the models of our targets and attempted an estimate of the star formation and BH properties.

The outline of this paper is the following: in \S 2 we present the two targets of our study, in \S 3 we present the X-ray \chandra observations, and in \S 4 we presented the ALMA archival data available for the two targets. In \S 5 we discuss our estimates of the properties for the two targets together with their implication for co-evolutionary models. Our conclusions are finally summarized in \S 6.

In this work, we adopt the standard flat cosmology (Planck Collaboration XIII 2016) with round parameter values:
matter density $\Omega_M = 0.32$, baryon density $\Omega_b = 0.05$, Hubble constant $H_0 = 100\,\hbox{h}\,\hbox{km}\,\hbox{s}^{-1}\,\hbox{Mpc}^{-1}$ with $h=0.67$, and mass variance $\sigma_8 = 0.83$ on a scale of $8\,\hbox{h}^{-1}\,$Mpc. Stellar masses and SFRs (or luminosities) of galaxies are evaluated assuming the Chabrier (2003) IMF.

\section{The targets: SDP.9 and SDP.11}\label{sec:targets}


SDP.9 (HATLAS J090740.0-004200) and SDP.11 (HATLAS J091043.1-000322) were firstly detected in the HATLAS first observations of a field $4\times4$ sq. deg. in size, taken during the Science Demonstration Phase (SDP, Maddox et al. 2010), that reached a 5$\sigma$ noise level of 33.5 mJy/beam at 250 micron (Negrello et al. 2010). They have a flux density above 100 mJy at 500 $\mu$m and are detected both in the SPIRE and PACS bands (i.e. between 100 to 500 $\mu$m, Eales et al. 2010, Poglitsh et al. 2010).

The Z-Spec spectrometer observed the lens candidates and detected CO lines at redshifts of z = 1.577$\pm$0.008 and z = 1.786$\pm$0.005 for SDP.9 and SDP.11 respectively (Lupu et al. 2012). Optical images of the foreground lens with the Keck telescope at g- and i-bands confirmed that their optical profiles are consistent with elliptical galaxies (Negrello et al. 2010).Follow-up observations with the SMA at 880 μm  (Bussmann et al. 2013) and with the Hubble Space Telescope  at 1.1 and 1.6 micron (Negrello et al. 2014) confirmed them to be lensed galaxies.

Dye et al. (2014) modeled the lensed systems as being lensed by elliptical galaxies at redhift $0.6129\pm0.00005$ and $0.7932\pm0.0012$. The magnification factors $\mu$ are estimated to be 6.29 and 7.89 (8.8 and 10.9 according to Bussmann et al. 2013) respectively for the two galaxies (see table 2 of Negrello et al. 2014 for a comprehensive summary of all the measures flux densities and modeled properties).

The selected objects have 12 and 22 $\mu$m flux densities, measured by the Wide-field Infrared Explorer (WISE; Wright et al. 2010) survey, in ‘excess’ by up to an order of magnitude to what expected from purely star-forming SED templates, thus suggesting the presence of a dust-obscured AGN. However, mid-IR excesses indicate that AGN, if present, are weak.
From the spectral decomposition made possible by the combination of the available data over a wide frequency range, Negrello et al. (2014, see their figure 6) evaluate the AGN bolometric luminosities $L_{\rm AGN}\sim5.1\times 10^{45} {\rm erg/s}$ for SDP.9 and  $L_{\rm AGN}\sim4.3\times 10^{46} {\rm erg/s}$ for SDP.11, although these values may be overestimated as they have been evaluated in a poorly sampled spectral region where contamination might enhance the signal in an unpredictable way. For these reasons we consider these values as a pure indication for the presence of an AGN and as loose upper limits of its luminosity.

In the near-IR SDP.9 consists of a dominant emitting region of $\sim1$ kpc in size, and a smaller and fainter one separated by a few kpc responsible for the fainter structure observed in the $\sim0.71$ arcsec radius Einstein ring (Dye et al. 2014). By the SED fitting, Negrello et al. (2014) estimated that the background source has a mass in stars $M\star \sim 7.1 \times 10^{10} M_{\sun}$. Lupu et al. (2012) estimated a comparable mass in molecular gas, $M_{\rm gas} \sim 3.4 \times 10^{10} M_{\sun}$ on the basis of CO emission. The gas fraction is $f_{\rm gas} = 0.32$, and the galaxy forms stars at a rate of $366\ M_{\sun}/yr$.

SDP.11 is reconstructed, in the near-IR, in a complex combination of several knots distributed within a region of a few kpc, which are responsible for the small-scale structure observed in the $\sim0.84$ arcsec radius Einstein ring (Dye et al. 2014). According to the SED fitting, compared to SDP.9, the source has a higher star formation rate, SFR$=650\ M_{\sun}/yr$, and a higher mass in stars $M\star \sim 1.9 \times 10^{11} M_{\sun}$. Lupu et al. (2012) CO observations indicate a comparable gas mass $M_{\rm gas} \sim 3.0 \times 10^{10} M_{\sun}$  and hence a lower gas fraction, $f_{\rm gas} = 0.14$.

We note that Lupu et al. (2012) adopted a linear relation ($M_{gas}=\alpha_{CO}L_{CO}$) to convert CO luminosity to gas mass with slope $\alpha_{CO}=0.8$, comparable with typical values observed in local ULIRGs. Recent observations in the radio and mm bands of starburst galaxies at $z> 1.5$ (Aravena et al. 2016, Bethermin et al. 2016, Popping et al. 2017) indicate that $0.8<\alpha_{CO}<1.5$ for populations of galaxies similar to our targets. This indicates that our mass estimates are most likely a lower limit of the actual gas mass content of the galaxies.

We also note that the SED fitting procedure implies uncertanties as high as $\sim50$ per cent of the stellar mass, and systematic errors can be due to the uncertain separation between lens and source and can provide up to an additional factor $\sim2$, but these are not crucial to our following findings.

Emission from the region of the two targets is present in the FIRST\footnote{http://sundog.stsci.edu/index.html} maps at 1.4 GHz at levels of $1.21\pm0.14$ and $0.72\pm0.12$mJy respectively for SDP.9 and SDP.11. The presence of emission from stronger sources within a few arcmin does not allow us to fully discard the possibility of a small spurios contamination, at least for SDP.9. The conversion of the SFR, as estimated form the Herschel maps, into radio signal at this frequency using the Bonzini et al. (2015) relation after scaling for the magnification factor does not discard the possibility of an AGN contribution responsible for the remaining emission.
   
\begin{figure}
   \centering
   \includegraphics[width=0.5\textwidth]{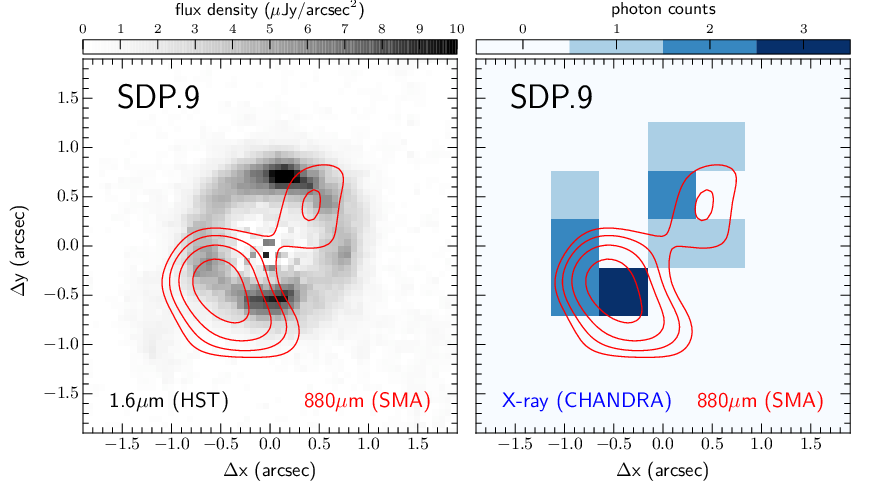}
   \includegraphics[width=0.5\textwidth]{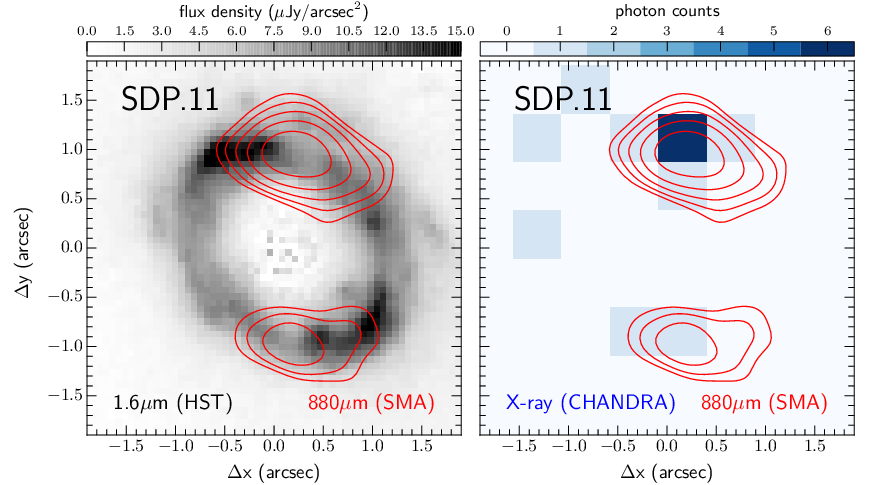}
   \caption{HST/WFC3 lens-subtracted F160W images (brightness scale in $\mu$Jy/arcsec$^2$, left panel) and \chandra 0.5/7keV images (right panel) of SDP.9 and SDP.11, overlaid with signal-to-noise ratio contours at 880 $\mu$m from the SMA (in steps of 3,6,9, Bussmann et al. 2013, Negrello et al. 2014).}          \label{fig:chandra_im}%
\end{figure}

\section{\chandra X-ray observations}
SDP.9 and SDP.11 were observed by \chandra in Cycle~16, with the ACIS-S3 CCD at the aimpoint, for a total exposure time of 55.3~ks (on January 10$^{th}$, 2015) and 19.8~ks (on February 28$^{th}$ and March 23$^{rd}$, 2015),
respectively.
Very-faint mode was used for the event telemetry format, and ASCA grade 0, 2, 3, 4 and 6 events were used in the analysis, which has been carried out using the {\sc ciao} v4.7 software.
Both sources are clearly detected, with 14 counts each in the 0.5--7~keV band within a circular region of 2\arcsec\ radius centered on the positions reported in $\S$2, and eight counts in the observed 2--7~keV band.
The 0.5--7 keV \chandra images of SDP.9 and SDP.11, centered at those positions, are shown in Fig.  \ref{fig:chandra_im}.
Remarkably, the 0.5 arcsec \chandra on axis resolution is able to resolve the X-ray emission from the lensed components. The X-ray emission shows a better positional match with the SMA data rather than with the HST images. 

The low number of detected photons limits significantly the spectral constraints that can be obtained from the X-ray analysis; however, to provide basic information for both sources, we fitted their spectra, extracted from the \chandra data, using simple models (within {\sc xspec} v.12.9, Arnaud et al. 1996) and the Cash statistics (Cash 1979), which is appropriate for low-count spectra.

Fitting SDP.9 spectrum with a powerlaw provides $\Gamma=1.1\pm{1.0}$, suggestive of obscuration. If we fix $\Gamma$ to 1.8, as typically observed in AGN and quasars (e.g., Piconcelli et al. 2005), we obtain a column density of $\approx5.2\times10^{22}$~cm$^{-2}$ at the source redshift (3$\sigma$ upper limit of $\approx3.3\times10^{23}$~cm$^{-2}$). The observed 2--10~keV flux is $\approx3.6\times10^{-15}$~\cgs, and the derived, apparent rest-frame 2--10~keV luminosity (corrected for the absorption but not for the magnification) is $\approx3.7\times10^{43}$~\lum.
The faint flux of SDP.9 explains why the source was undetected (with an upper limit of $\approx10^{-14}$~\cgs\ in the 0.5--8~keV band) in the XMM-ATLAS survey (Ranalli et al. 2015), that covers a 6~deg$^2$ area with a 3~ks average exposure. The non-detection of the AGN in \xmm\ is suggestive of the presence of obscuration, with an estimated column density of at least $10^{23}$~cm$^{-2}$, as observed in other bright submm galaxies (Johnson et al. 2013).

For SDP.11, we combined the photons of the two observations into a single spectrum. As for SDP.9, a powerlaw model provides an apparently flat spectrum, with $\Gamma=1.0\pm{0.9}$. If $\Gamma=1.8$ was assumed, the column density would be $3.2\times10^{22}$~cm$^{-2}$ (3$\sigma$ upper limit of $\approx2.4\times10^{23}$~cm$^{-2}$).
The observed 2--10~keV flux is $\approx1.2\times10^{-14}$~\cgs, and the apparent rest-frame 2--10~keV luminosity (again corrected for the absorption but not for the magnification) is $\approx1.2\times10^{44}$~\lum.

\begin{figure}
\centering
\includegraphics[width=0.5\textwidth]{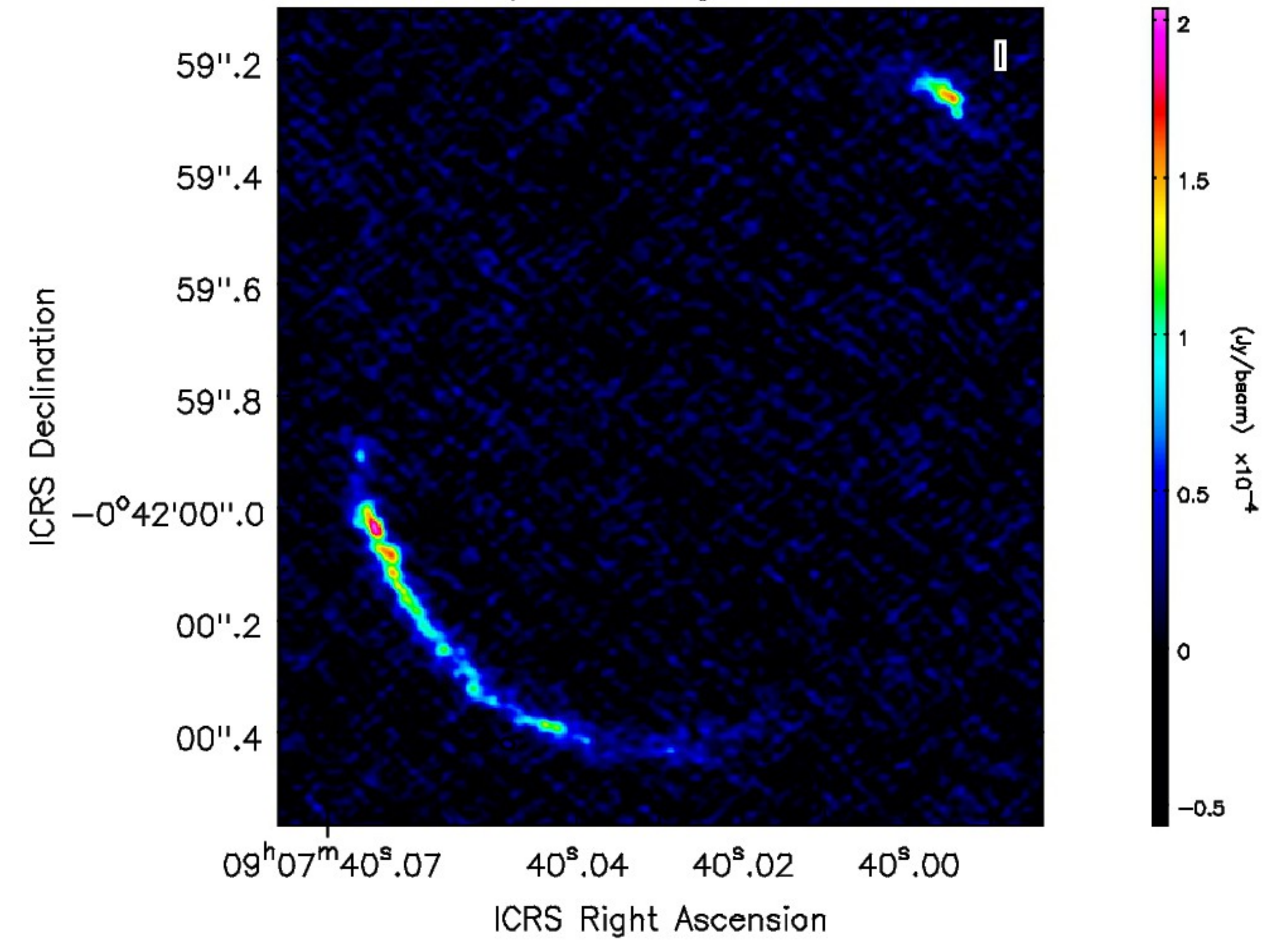}
\caption{ALMA high resolution continuum 1.3mm observations of SDP9.}
         \label{fig:alma_continuum}
\end{figure}

  \begin{figure}
   \centering
   \includegraphics[width=9cm]{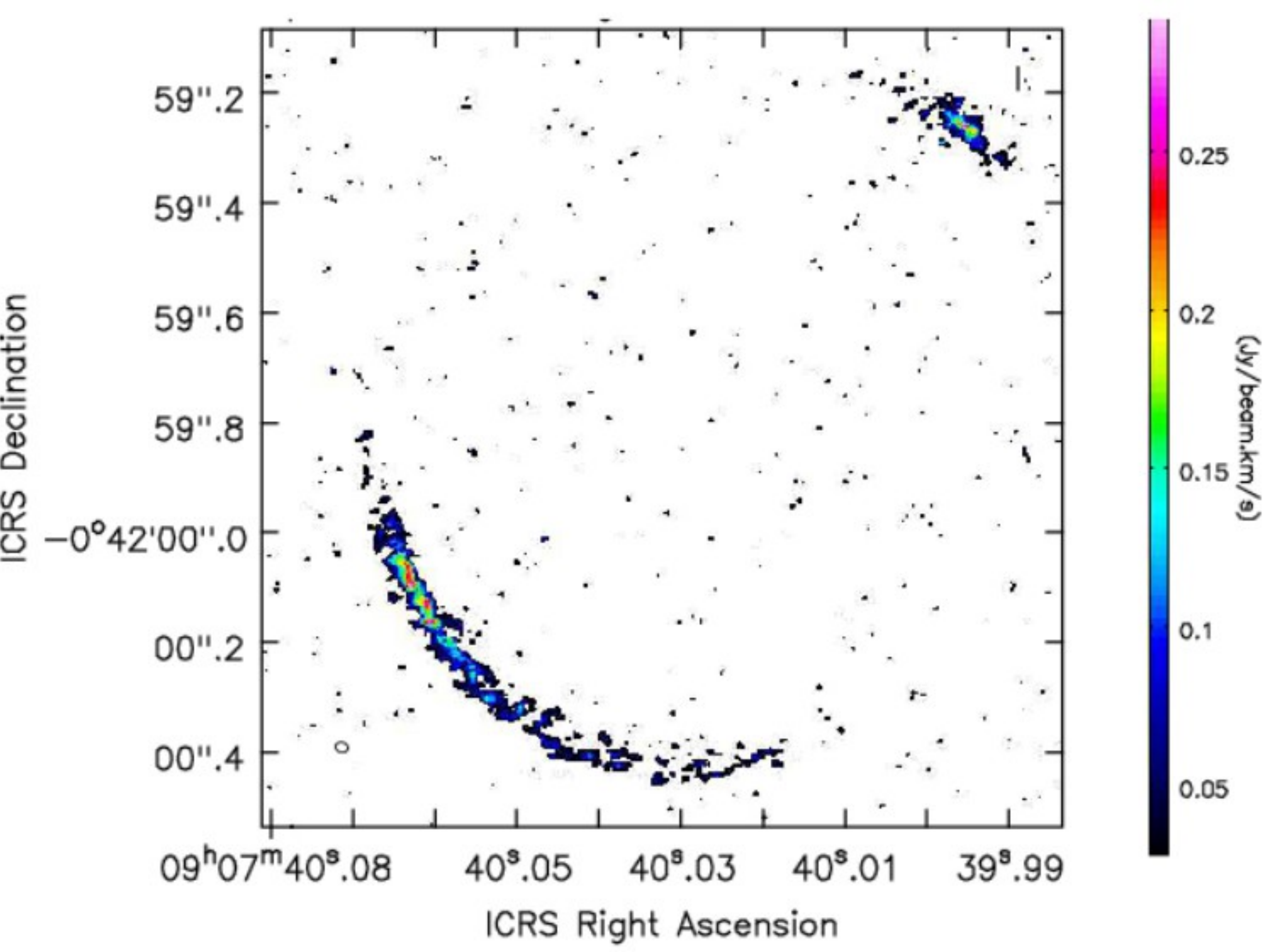}
   \includegraphics[width=9cm]{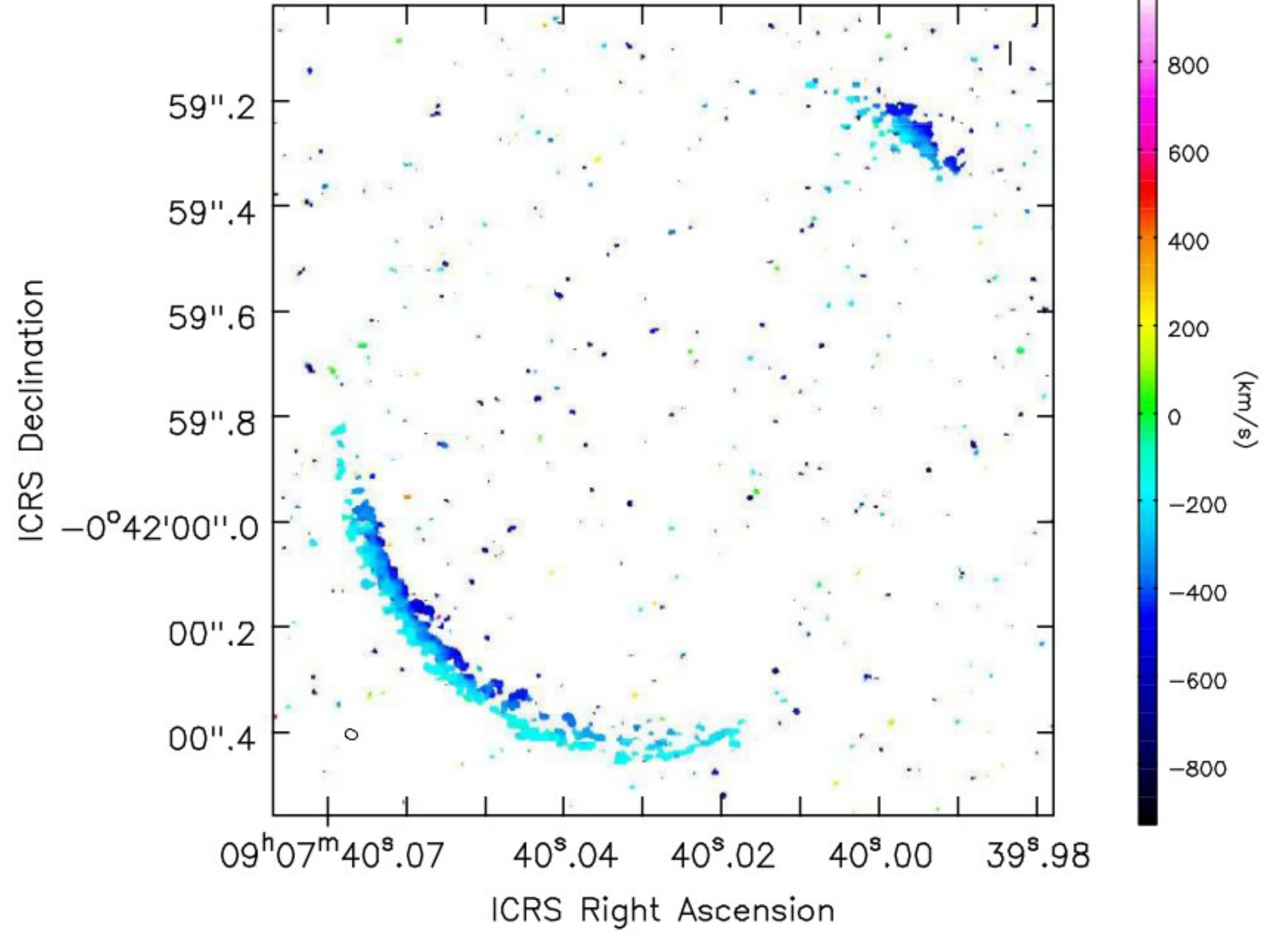}
   \includegraphics[width=9cm]{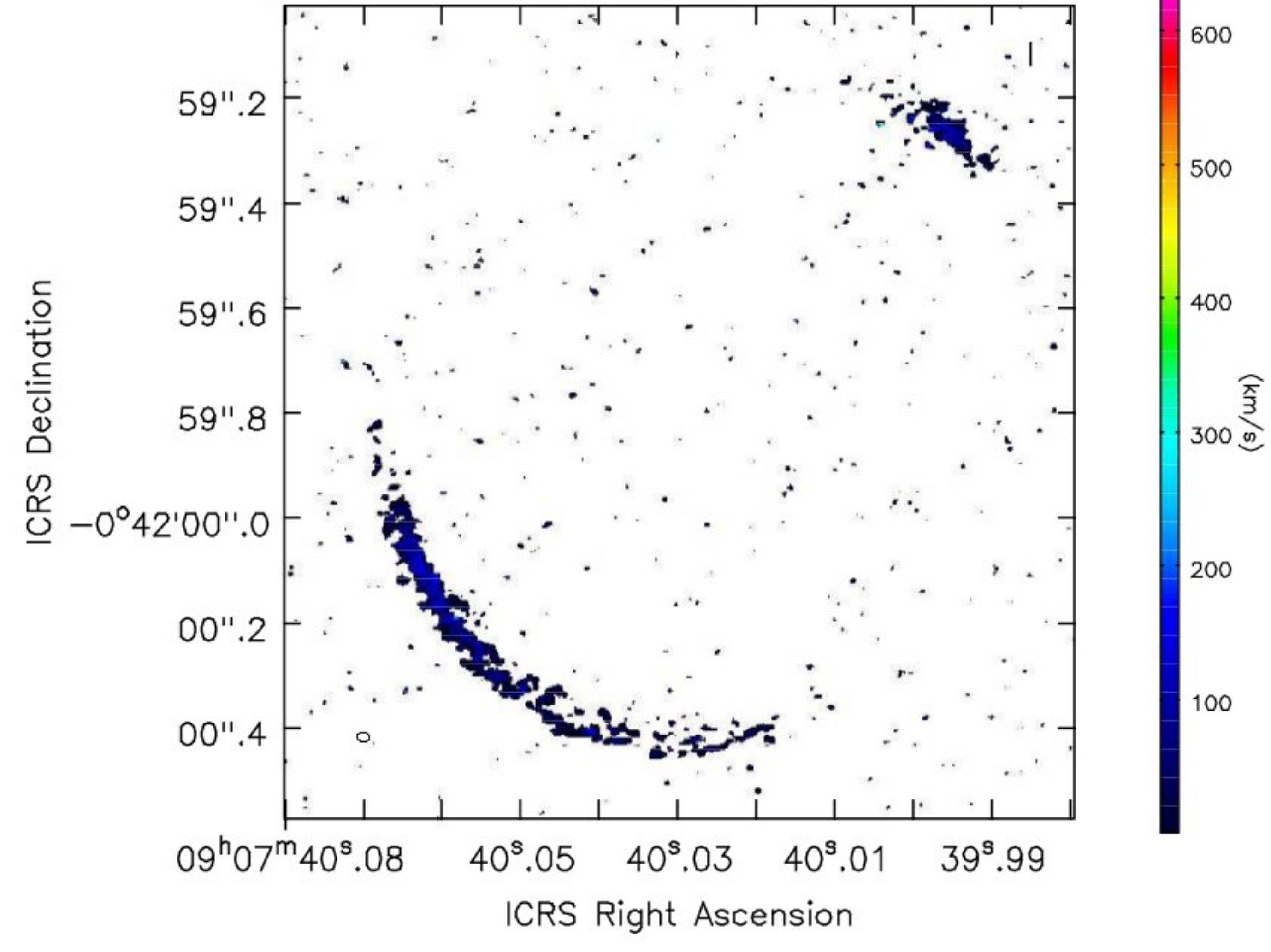}
   \caption{SDP9 CO(6-5) 0, 1, and 2 momenta of the line distribution, corresponding to the integrated brightness, and velocity distributions and the velocity dispersion.}
              \label{fig:alma_momenta}%
    \end{figure}

Once corrected for the magnification factors, the intrinsic X-ray luminosities are $\approx5.9\times10^{42}$~\lum\ and $\approx1.5\times10^{43}$~\lum\ for SDP.9 and SDP.11, respectively.
We can estimate the contribution from star-formation processes to the 2--10~keV luminosities assuming the relation reported in Ranalli et al. (2003) and the SFRs obtained from SED fitting ($\S$2).
The derived values are $\approx1.9\times10^{42}$~\lum\ (SDP.9) and $\approx3.3\times10^{42}$~\lum\ (SDP.11), i.e., the X-ray intrinsic luminosities are a factor $\approx$3--4.5 higher than those possibly associated with stellar processes. Recently, based on the analysis of the 6Ms CDFS data, Lehmer et al. (2016) have derived more refined scaling relations between the X-ray luminosity of galaxies and their SFR, stellar mass and redshift of the form $L_{2-10keV}=\alpha M_\star+\beta SFR$, where $\alpha$ and $\beta$ increase with redshift. By applying the scaling relations of Lehmer et al. (2016; see their Table~4), the 2-10 keV
luminosities that can be ascribed to stellar processes would be $\sim30\%$ higher than those estimated with the Ranalli et al. relation. We note that because of the large, $\approx 30-40\%$ uncertainties in the measured X-ray luminosities, dominated by the low photon statistics,
the measured values would still be compatible within $\sim2.0-2.5\sigma$ with being produced by star formation. However, the systematically higher measured values point again towards an AGN-related nature.

We finally note that any X-ray emission associated with star formation should occur primarily at low energies. This would produce an apparently lower value for the measured nuclear obscuration in low-counting statistics spectra as those of SDP.9 and SDP.11, making the intrinsic X-ray luminosities quoted for the two objects likely lower limits.

   \begin{figure*}
   \centering
   \includegraphics[width=\textwidth]{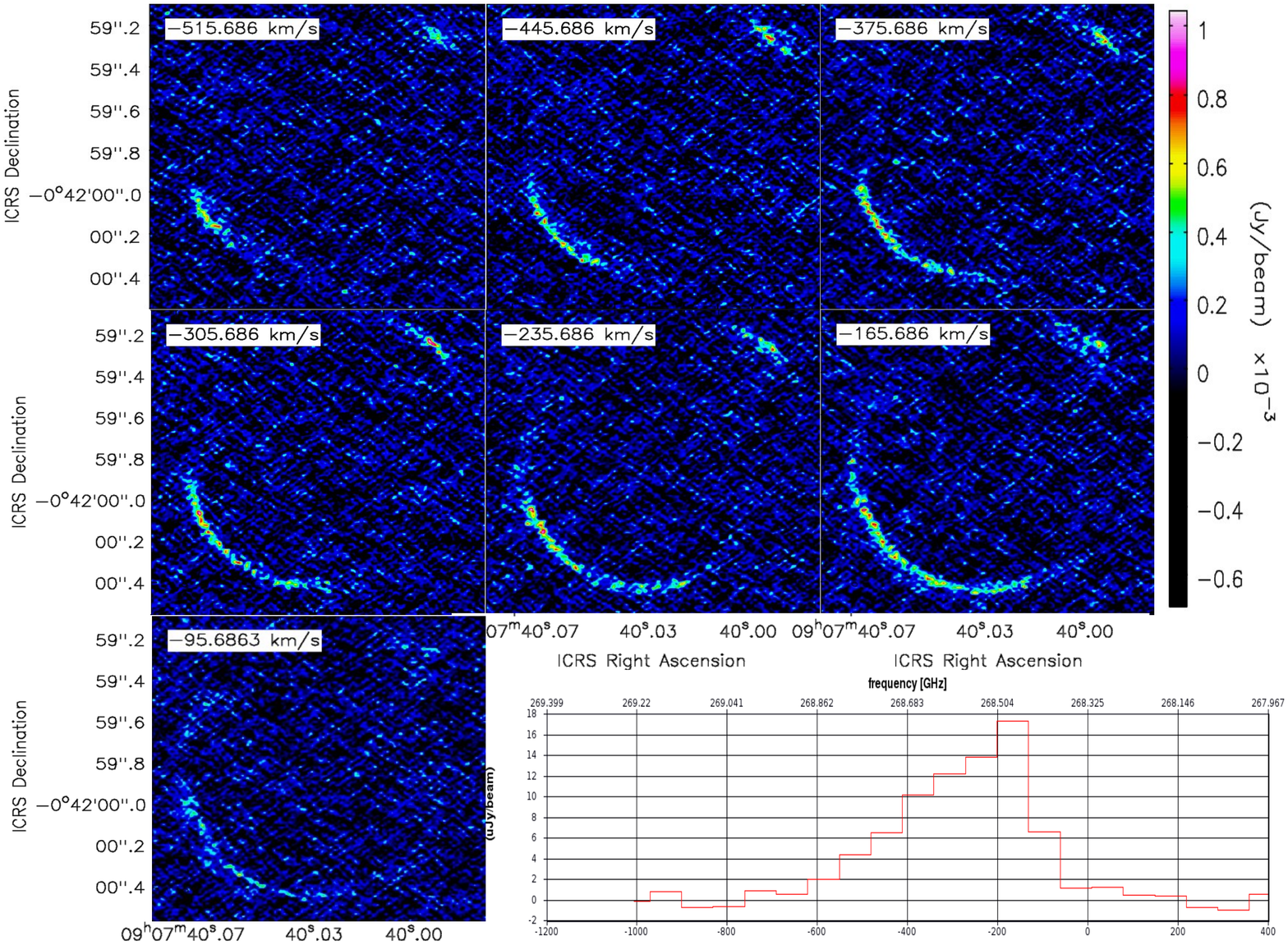}
   \caption{SDP9 CO(6-5) channel maps and spectrum of the whole galaxy. The 0 km/s was assumed corresponding to the expected position for CO(6-5) redshifted according to the past estimates by Bussmann et al. (2013). }
              \label{fig:alma_CO}%
    \end{figure*}

\begin{figure*}
   \centering
   \includegraphics[width=0.47\textwidth]{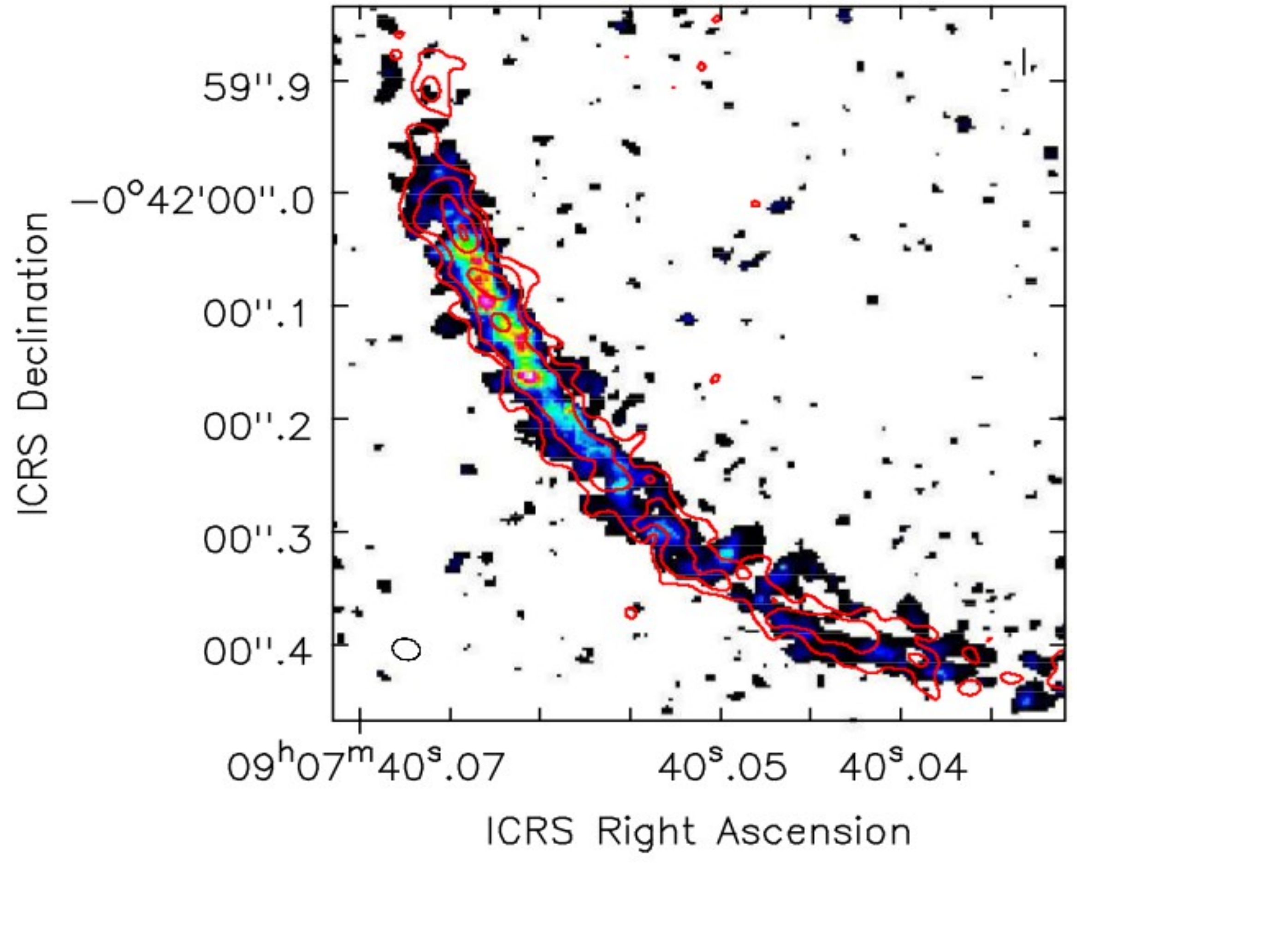}
   \includegraphics[width=0.47\textwidth]{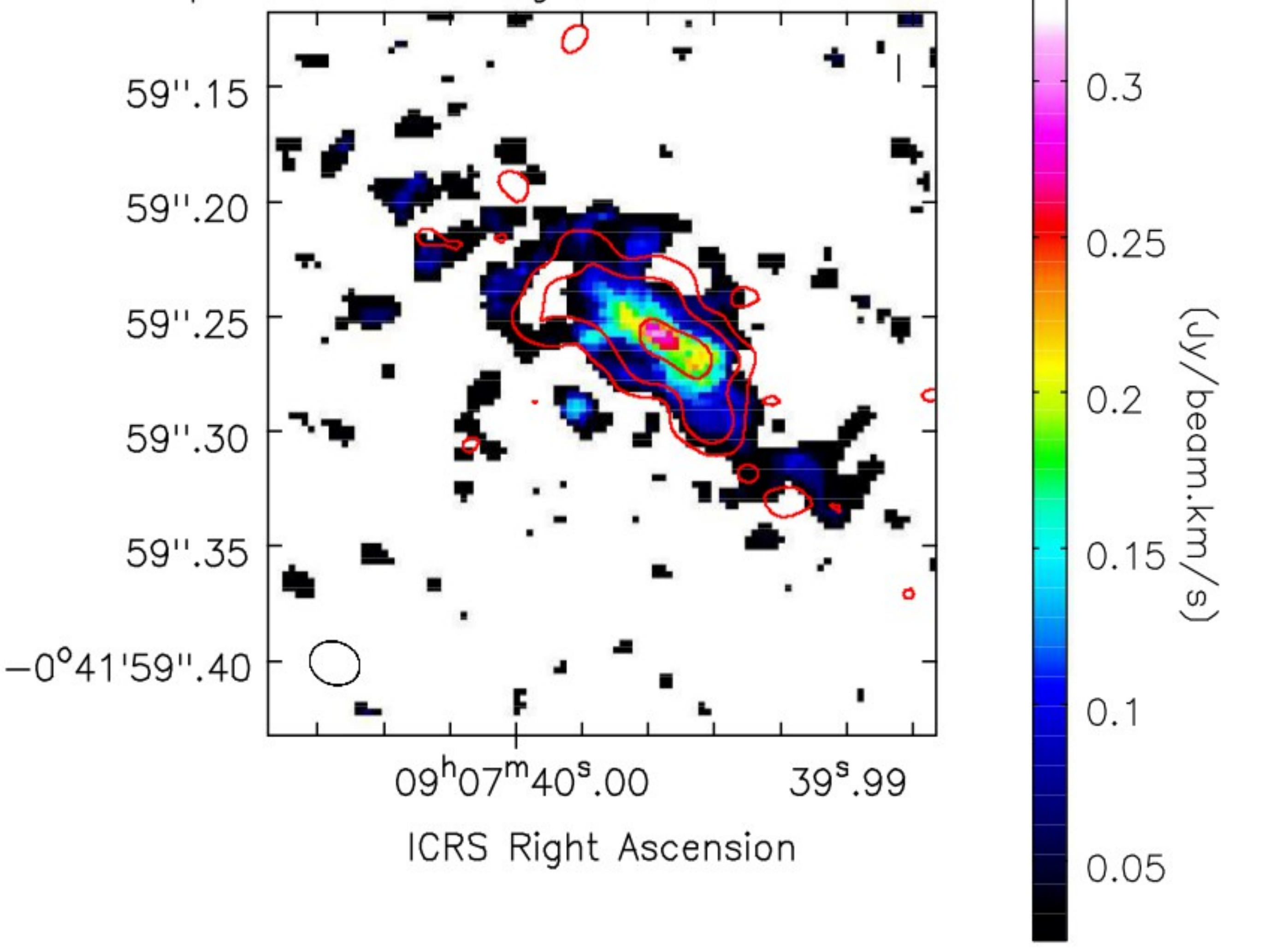}
   \caption{SDP9 CO(6-5) brightness distribution overlaid with the continuum contours (6,12,18 $\sigma$) for the NW and the SE images of the lensed galaxy.}
              \label{fig:alma_COandcont}%
    \end{figure*}

\section{ALMA mm observations}

The two targets were included in a low-resolution observation with ALMA in Cycle 1 (2012.1.00915, PI:Lupu). On the basis of those data, Oteo et al. (2017) investigate dense circumnuclear region emissions through HCN(3-2), HCO$^+$(3-2), HNC(3-2) and CO(3-2) transitions properties. HCN(3-2), HCO$^+$(3-2) and CO(3-2) lines have been detected with ALMA for both the targets while HNC(3-2) was detected only in SDP.9. These spectral line intensity ratios and their relation with the continuum emission suggest that the presence of intense star formation could be associated with massive dense molecular gas reservoirs and high dense molecular gas fractions.

High resolution ($<0.1$ arcsec) observations have been carried out for both our targets. The observations are still on going for the SDP.11 (2015.1.01362.S, PI:Gordon). In the following we report our reduction, imaging, analysis of the data for SDP.9 available in the ALMA Science archive of Cycle 4 project, performed with a different approach with respect to those recently presented by Wong et al. (2017). Presenting our detailed image analysis is necessary as they are the input to the modeling of SDP.9 that will be discussed in the next section.

\subsection{SDP 9 high resolution observations}
ALMA Cycle 3 observations for SDP9 were obtained in the framework of a project aiming at detecting with subarcsec resolution traces of the emission from the foreground lens (ID: 2015.1.00415 PI: Wong).
A 1.875GHz-wide spectral window was centered at 268.34GHz to include the CO(6-5) emission redshifted according to the 1.577$\pm0.008$ value estimated in Bussmann et al. (2013). Three further spectral windows centered at 266.21, 254.01 and 251.51 GHz were used for continuum estimation. The native spectral resolution of the data is 31.250MHz (corresponding to $\sim35$km/s after Hanning smoothing).

Baselines as long as 16.196 km were used to reach a 0.02 arcsec resolution.
A total of 4.3h, split in two execution blocks, was enough to reach 0.1$\mu$Jy sensitivity over the $\sim7.3$ GHz band used for the continuum (after flagging the channels including the CO(6-5) line).

Calibration was done manually and no major issue was found. Imaging was performed in an interactive way to apply a mask, with standard Briggs weighting scheme (robustness parameter set to 0.5) and no tapering.
The final maps have 0.02$\times$ 0.02 arcsec resolution (corresponding to $\sim65$pc on the lensed galaxy plane). 

The continuum peak was detected to $>20\sigma$ (see fig. \ref{fig:alma_continuum}).
The CO(6-5) image cube was generated with a 70km/s spectral resolution (see fig. \ref{fig:alma_CO}) to increase sensitivity on the line tail. The peak of the spectrum was offset with respect to the expected position, and with an asymmetrical tail towards higher frequencies. The peak estimated over a region that includes the whole galaxy corresponds to a redshift 1.5753$\pm0.0003$, assuming the FWHM of the tail on the higher redshift, that has a Gaussian profile as an estimation of the error, consistent with the past estimates (1.577$\pm0.008$), but evaluated now to a better accuracy. The value of 1.5747$\pm$0.0002 reported by Wong et al. (2017) is well within the line size, but we caveat that both ours and their quoted accuracy values do not take in consideration the asymmetric tail of the line that extends more than $250\ $MHz (half width at half maximum), corresponding to a $\Delta z \sim 0.0024$, improving the accuracy quoted by Bussmann et al. (2013).

While the overall distribution of continuum and CO brightness overlap along the ring (see fig. \ref{fig:alma_COandcont}), the velocity distribution pattern (see the second panel in fig. \ref{fig:alma_momenta}) shows a clear geometrical offset in the velocity components that could be associated with two separated emitting components. In some of the knots a bimodal distribution of the line is clearly visible: in such cases the difference in velocity among the two components is $\sim300$km/s.     
\begin{figure*}
\begin{centering}
\includegraphics[width=\textwidth]{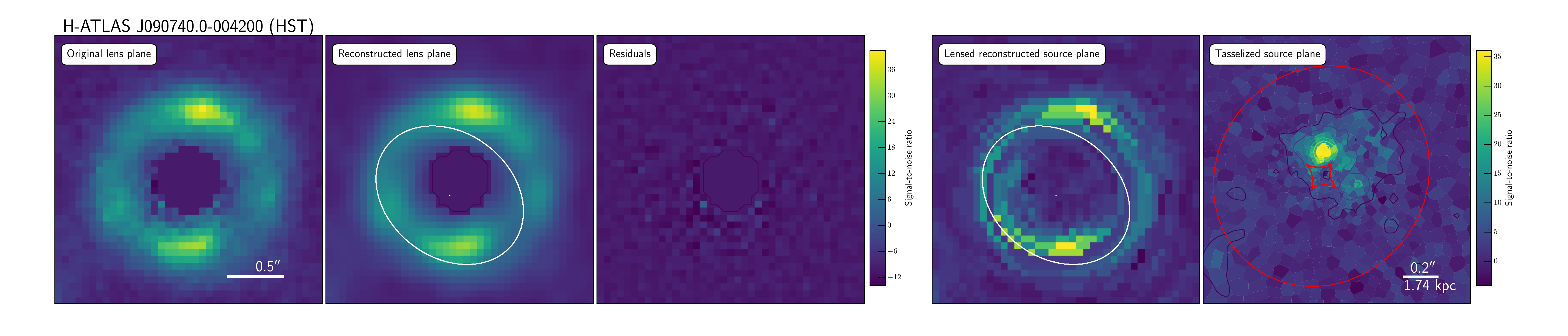}
\includegraphics[width=\textwidth]{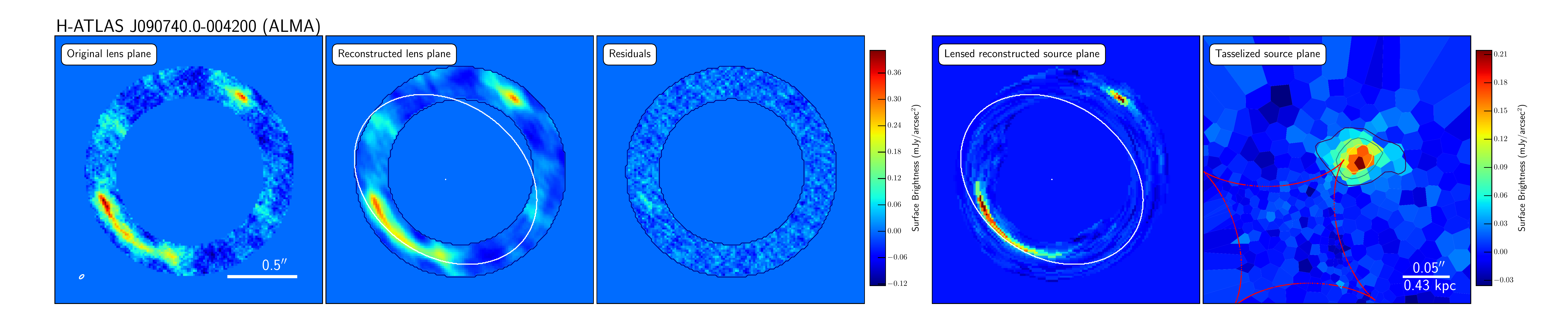}
\end{centering}
\caption{Lens modeling results on the HST/WFC3 data (top) and ALMA (bottom). Left panels: the original lens plane. The tiny white ellipse at the
bottom left of the ALMA original lens plane is the final image restored beam. The best-fit lens plane is shown in the central panel of the left column; the panel on its right-hand shows the residuals. Right panels: the reconstructed source plane on Voronoi tassels. Contours are from 3$\sigma$ level to the maximum S/N with steps of 3 (ALMA) and 5 (HST). White and red curves are the lens caustics and critics respectively on the
lens plane and on the source plane; angular and physical scale is reported in white on the bottom-right of every picture. 
\label{fig:lens_modeling_results}}
\end{figure*}
    
\begin{table*}
\centering
\caption{Lens modeling results for the HST 1.6 $\mu$m and ALMA 1.3 mm data for SDP9. Uncertainties are $\pm$ 1$\sigma$. Einstein radii are in arcseconds, angles are in degrees measured counter-clockwise from East. Lens positions are referred to the center of the HST observation.}
\label{table:parameters}
\scriptsize
\begin{tabular}{lccccc}
\hline
				& $r_{\rm E}$	& $\phi$			& $q$ 			& $\Delta x_{\rm L}$ & $\Delta y_{\rm L}$	\\
		 		& ''			& $\circ$			&				& ''				 & ''					\\
\hline
HST	 $1.6\mu$m	& 0.66$\pm$0.02 & 142.51$\pm$7.51	& 0.77$\pm$0.02	& -0.09$\pm$ 0.02	& -0.04$\pm$ 0.05		\\
ALMA $1.3$ mm	& 0.65$\pm$0.04 & 143.19$\pm$9.09	& 0.77$\pm$0.04	& -0.10$\pm$ 0.03	& +0.01$\pm$ 0.03 		\\
\hline
\end{tabular}
\end{table*}

\begin{table*}
\centering
\caption{Source properties for SDP9: magnification factors $\mu$, brightness area of the reconstructed source, and source radii within 3 and 5 $\sigma$ . Uncertainties are $\pm$ 1$\sigma$.}
\label{table:sourceprop}
\scriptsize
\begin{tabular}{lcccccc}
\hline
		 		& $\mu_{3\sigma}$ & $\mu_{5\sigma}$ & $A_{3\sigma}$	& $A_{5\sigma}$	& $r_{3\sigma}$		& $r_{5\sigma}$		\\
				& 				        & 			        	& kpc$^2$ 		  & kpc$^2$   		& kpc 				    & kpc 				\\
\hline
HST	 $1.6\mu$m	&  7.80$\pm$0.44  &  8.32$\pm$0.49	& 20.43$\pm$1.8	& 11.45$\pm$1.6	& 2.550$\pm$0.117	& 1.909$\pm$0.144	\\
ALMA $1.3$ mm	& 17.39$\pm$3.86  & 18.73$\pm$4.43	& 0.82$\pm$0.34&0.44$\pm$0.16 &0.510$\pm$0.098	& 0.375$\pm$0.064	\\
\hline
\end{tabular}
\end{table*}

\begin{figure*}
\begin{centering}
\includegraphics[width=\textwidth]{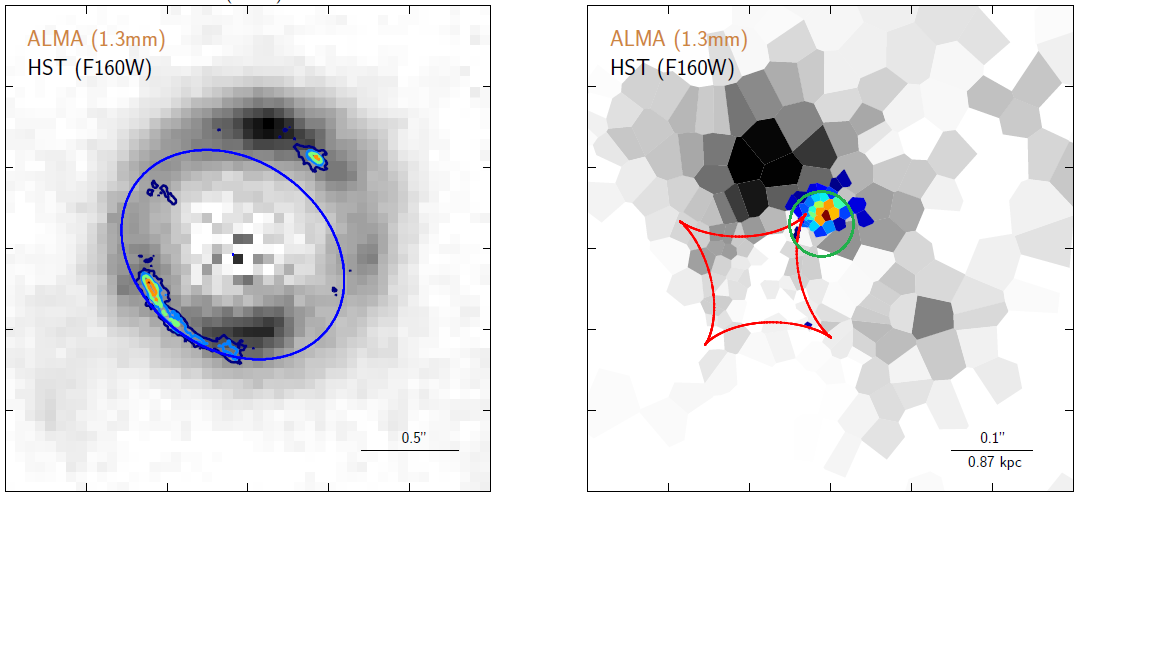}
\end{centering}
\caption{Left: the observed original lens plane. Right: the reconstructed source plane. Grey scale
is the HST/WFC3 1.6 $\mu$m data, rainbow colorscale the 1.3 mm ALMA data. The green circle marks the position of the reconstructed \chandra peak within $5\sigma$.}
\label{fig:composite_results}
\end{figure*}

\section{Modeling the emission of SDP9}

We perform lens modeling, in order to obtain the lens parameters, reconstruct the source emission and retrieve other useful informations such
as the total magnification $\mu$. There are mainly two different approaches to lens modeling. 
The so called `parametric' methods describe both the lens and the source profile with analytical models (such as the Singular Isothermal Ellipsoid or the Sersic profile), and they explore the parameter space to obtain the set that minimizes a $\chi^2$ statistics between the observed and modeled surface brightness counts distribution on the lens plane. Whenever the source morphology is complex or clumpy (such as in case of dusty star-forming galaxies) the results tend to be a oversemplification of the real physical situation. 
Non parametric' or 'semi-parametric' methods aim to reconstruct the original source morphology by assuming a pixellated distribution for the surface brightness of the background galaxy. They exploits a simple matrix formalism   with the introduction of a regularisation term that ensures a certain degree of smoothness in the reconstructed source (Warren and Dye 2003; Suyu et al. 2006)

In this work, we adopt the Regularized Semilinear Inversion method of Warren and Dye (2003), improved with the adoption of an adaptive source plane pixelization scheme (Nightingale and Dye 2015), and extended to deal with interferometric data (Dye et al. subm., Enia et al., in prep.). We refer the reader to these references for more details.

In order to reduce computational time, not every pixel enters the modeling alghoritm, only the ones inside a mask containing the lensed image, with minimun background sky. The lens is a Singular Isothermal Ellipsoid (SIE, Kormann et al. 1994), described with five parameters: the Einstein radius $r_{\rm E}$, the position angle $\phi$ defined as the orientation of the semi-major axis measured counter-clockwise from West, the ellipticity defined as the ratio between the minor and major axis of the ellipsoid, the displacement of lens centroid $\Delta x_{\rm L}$ and $\Delta y_{\rm L}$ with respect to the HST image center. The lens parameter space is explored using {\sc multinest} (Feroz et al. 2009).

Lens modeling results are reported in Tab. \ref{table:parameters}. Both data sets lead to the same set of parameters, which are similar to other results showed in literature (Bussmann et al. 2013, Dye et al. 2014), proving the strenght and goodness of our model.

The results are showed in Fig. \ref{fig:lens_modeling_results}. Properties such as magnification and sizes are computed perturbing 1000 times the lens model parameters around their best-fit values; the reported final values and errors are being the median and the 16th and 84th percentile of the distribution. For each iteration, magnifications are evaluated as $\mu=F^{\rm LP}_{3\sigma}/F^{\rm SP}_{3\sigma}$, $F^{\rm SP}_{3\sigma}$ being the total flux inside the source plane (SP) pixels with SNR $\geq3$ and $F^{\rm LP}_{3\sigma}$ the total flux inside the corresponding lensed pixels. Similarly, $r_{\rm eff}$ are measured from the reconstructed source area recalling that $A^{\rm SP} = \pi r_{\rm eff}^2$.

The retrieved magnification are respectively 7.80 and 17.39 for the HST/WFC3 and ALMA observations. The first value is higher than (but consistent within $2.5\sigma$ with) the value $μ_{tot}\sim6.3$ reported in Dye et al. (2014, but the lens was modelled with a Singular Power Law Ellipsoid and resulted to be more spherical than what we find), the second is almost twice the value of $\mu_{880}$ reported in Bussmann et al. (2013), even though it is worth noticing that in that case the lens modeling method was a fully parametric one, with the reconstructed source modeled with a Sersic profile. Anyway, the high value of $\mu_{1.3}$ is expected since the reconstructed source is compact and peaks close to the lens caustic. Source properties are reported in Table \ref{table:sourceprop}.

The observed lens planes and reconstructed source planes are superimposed in Fig. \ref{fig:composite_results}. From a visual inspection of the lensing morphology, there is a clear presence of a displacement between the two data sets. This is frequently observed in this kind of sources (Dye et al. 2015, Fu et al. 2012), and reflects the intrinsic nature of the observations. The near-IR emission traces the direct light of stars that are partially or not at all obscured by dust, while the sub-mm emission traces directly the dust-reprocessed photons from the UV radiation field of newborn stars, that is completely or almost completely obscured. This displacement is then reproduced on the reconstructed source plane, with a distance between the two emission peaks of $\sim0.9$ kpc.

A first order lens modeling with the previously found HST parameter set has also been performed on available \chandra X-ray data, in order to simply constrain the position of the peak signal, since any further analysis needs data at higher resolution. This is cospatial with the sub-mm signal, and along with the mid-IR excess observed in the source SED, suggests the presence of an obscured AGN in the galaxy.

\begin{figure*}
\includegraphics[height=6cm, width=0.45\textwidth]{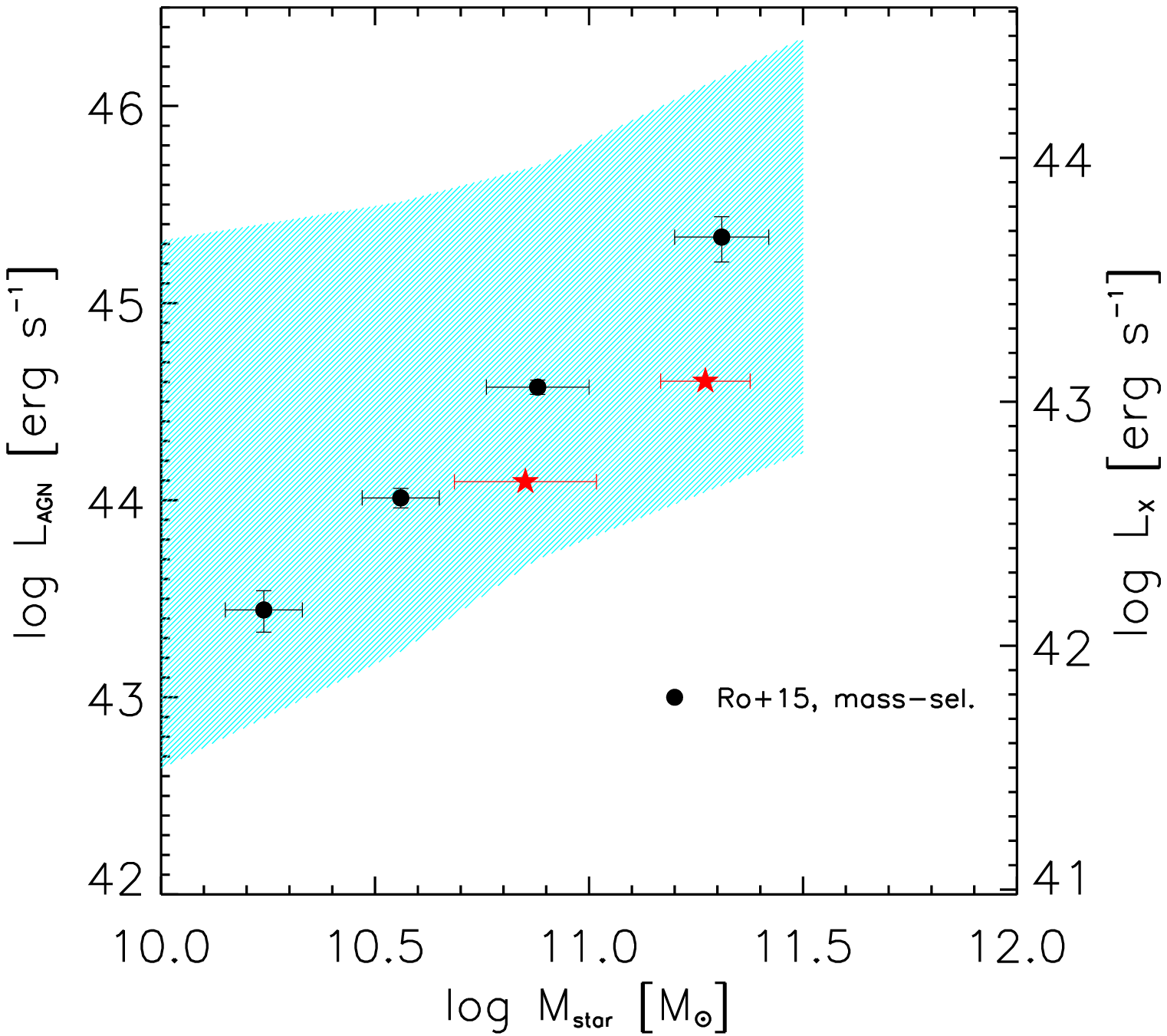}
\includegraphics[width=0.5\textwidth]{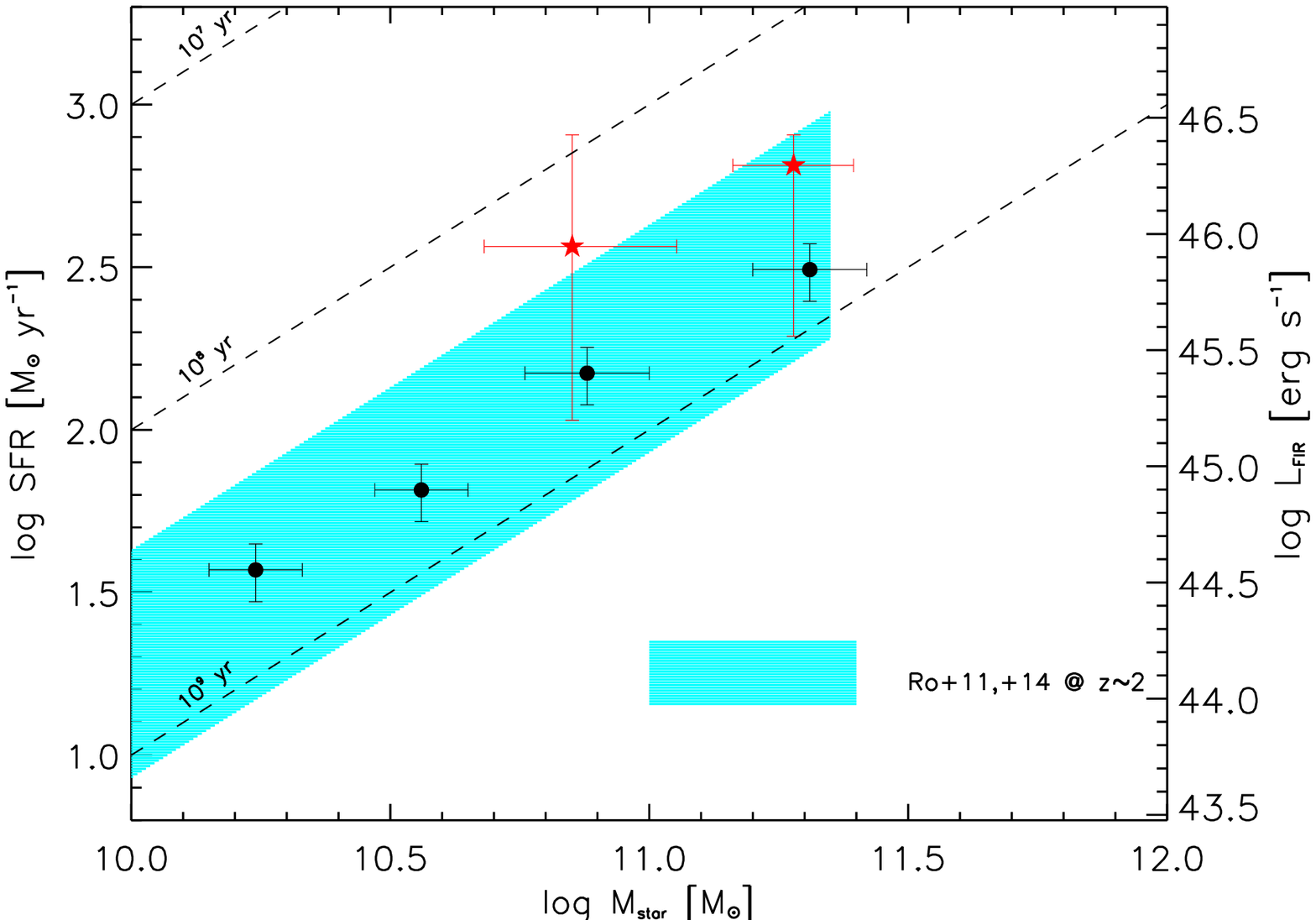}
\caption{The positions (red stars) of SDP9 and SDP11 in the main sequences of AGNs ($L_X$ or $L_{\rm AGN}$ vs. $M_\star$, left panel) and starforming galaxies (SFR vs. $M_\star$, right panel). Data based on a mass-selected sample are from Rodighiero et al. (2015; black circles), with the cyan shaded area illustrating the typical scatter around the mean relationship.}\label{fig:main_seq}
\end{figure*}

\section{Discussion}

Our \chandra-ACIS observations confirmed that by observing strongly-lensed submm-bright galaxies we can access earlier stages of the co-evolutionary scenario than previously possible. So far, in fact, available X-ray data allowed to test the in-situ coevolution scenario only at relatively old galactic ages $\la 10^9$ yr, when the black hole has grown enough for its X-ray luminosity to be detectable at substantial redshift ($z>1$).
This constitute a successful pilot study of the coevolution between black hole and galaxy growth at high redshifts, in a low X-ray luminosity regime associated to the early stages of galaxy evolution.

The two strongly-lensed galaxies analyzed in this work feature (intrinsic) SFRs of several $10^2\,M_\odot$ yr$^{-1}$ corresponding to far-IR luminosities $L_{\rm FIR}\ga 10^{46}$ erg s$^{-1}$; the stellar masses $M_\star$ are $\sim10^{11}\, M_\odot$, implying (for nearly constant SFR, see Papovich et al. 2011, Smit et al. 2012, Citro et al. 2016) rather young ages  $\tau_{\rm age}\approx M_\star/$SFR of a few $10^8$ yr. As such, in the SFR vs. $M_\star$ diagram at $z\sim 2$, these galaxies are slightly offset with respect to the main sequence locus (Rodighiero et al. 2011, 2014; Speagle et al. 2014; see Fig.\ref{fig:main_seq}). This is due to the fact that the main sequence locus has been classically defined for mass-selected samples, while the galaxies considered here are far-IR selected; mass selection tends to pick up starforming galaxies at the end of their main star formation period, when their age is $\tau_{\rm age}\la 1$ Gyr and their mass fully accumulated, while far-IR selection tends to pick up them unbiasedly with respect to age and stellar mass (see Mancuso et al. 2016b). Hence, our targets are intrinsically richer in gas because they are typically younger than main sequence galaxies with the same SFR. However, in the selection process we picked up object with specific SFR lower than other selected FIR samples in order to have higher chance of observing the X-ray nuclear emission.

In fact, SDP.9 and SDP.11 still retain appreciable amounts of gas $M_{\rm gas}\ga a few 10^{10}\, M_\odot$, implying depletion timescales $\tau_{\rm depl}\approx M_{\rm gas}/$SFR$\ga$ several $10^7$ yr. Provided that further gas infall may well occur from a hot atmosphere, this is actually a lower limit to the depletion time; the star formation will likely proceed for a more realistic $\tau_{\rm depl}\ga$ a few $10^{8}$ yr, when the objects will then move to lie onto the main sequence. Furthermore, a higher value for the $\alpha_{CO}$ conversion factor, could enhance the gas mass by a factor of $\lesssim 2$ and, as a consequence allow larger values of the depletion time by similar factors.

As to the nuclear emission, the intrinsic (demagnified and obscuration corrected) X-ray luminosities amount to $L_X\la 10^{43}$ erg s$^{-1}$; note that these values are appreciably above (factor $3-4.5$) the expected X-ray emission from star formation $L_{X,{\rm SFR}}\ga 10^{42}$ erg s$^{-1}$. The nuclear bolometric luminosities (using the Hopkins et al. 2007 corrections) are of order $L_{\rm AGN}\ga 10^{44}$ erg s$^{-1}$, a factor $10^{2}$ smaller than $L_{\rm FIR}$ and corresponding to a BH mass $M_{\rm BH}\approx 10^6\, M_\odot$, for an Eddington ratio around unity (as expected in these early stages when plenty of material is available for accretion, Alexander et al. 2008). As  such the two galaxies are located slightly below the main sequence of AGNs (Mullaney et al. 2012;  Rodighiero et al. 2015; see Fig. 2). Adopting the standard ratio $M_{\rm BH}/M_\star\approx$ a few $10^{-3}$ between the final BH and stellar masses would imply a relic BH mass of $M_{\rm BH}\approx 10^{-3}\times 10^{11}\, M_\odot\ga 10^8\, M_\odot$. The corresponding bolometric luminosities will be around $L_{\rm AGN}\approx 10^{46}$ erg s$^{-1}$, comparable to the far-IR luminosity from star formation; at that time, plausibly the energetic feedback from the AGN will quench the star formation  process and terminate further accretion onto the BH.

In order to increase its mass by a factor $10^2$ from the observed to the expected final value over $\tau_{\rm depl}\approx$ a few $10^8$ yr, the BH must grow in a self-regulated, exponential fashion; indeed the related accretion $e-$folding timescale $\tau_{\rm ef}\, \ln 10^2\approx 5\, \tau_{\rm ef}\approx$ a few $10^8$ yr is found to be consistent with $\tau_{\rm depl}$. This implies that: (i) star formation and BH accretion are not always proportional to each other during galaxy evolution; (ii) the most likely mechanism to buildup the final BH-to-stellar-mass ratio should allow low-angular momentum gas to be stored in a massive circumnuclear reservoir at a rate proportional to the SFR. For the two galaxies discussed here, a reservoir mass of order $10^8\, M_\odot$ is implied; the size $R_{\rm res}$ of the reservoir should correspond to several times the BH influence radius $G\, M_{\rm BH}/\sigma^2\approx R_e\, M_{\rm BH}/M_\star$, and thus should amount to $R_{\rm res}\approx 10-100$ pc for an effective radius $R_e\approx 10$ kpc.

A crucial probe of the above interpretation will be provided by observations of the reservoir in strongly lensed, dust enshrouded galaxies. In fact, strong gravitational lensing with amplification factors $\approx 10$ can bring the apparent size of a reservoir with physical size $R_{\rm res}\approx 10-100$ pc to $\approx 0.02-0.07$ arcsec, well within the resolution achievable with ALMA (see Kawakatu et al. 2007; Maiolino 2008; Spaans \& Meijerink 2008; Lapi et al. 2014). Our modeling of SDP.9 strongly indicates that dust and X-ray emission coexist in an unresolved region of size smaller than 375 pc (see table \ref{table:sourceprop} and figure \ref{fig:composite_results}). 

\section{Conclusions}

We presented the \chandra and ALMA observations of two strongly lensed galaxies, SDP.9 (HATLAS J090740.0-004200) and SDP.11 (HATLAS J091043.1-000322) that we selected in the Herschel-ATLAS catalogues as having an excess emission in the mid-IR regime and relatively low specific star formation rate with respect to common FIR selected samples at redshift $\gtrsim1.5$, indicating the possible presence of a nuclear activity in the early stages of the galaxy formation.
Both the targets were detected in the X-ray, strongly indicating the presence of highly obscured nuclear activity. 
   
ALMA observations for SDP9 for continuum and CO(6-5) spectral line with high resolution (0.02arcsec corresponding to $\sim65$ pc at galaxy distance) allowed us to estimate the lensed galaxy redshift to a better accuracy than pre-ALMA estimates ($1.5753\pm0.0003$) and to model the emission of the optical, millimetric, and X-ray band emission for this galaxy. We demonstrated that both the millimeric and the X-ray emission are generated in the nuclear environment, thus strongly supporting the presence of nuclear activity in this object. The X-ray model, despite its low resolution ($\sim0.5$ arcsec), clearly peaks in the nuclear region, in strong concordance with the mm peaks. This confirms that the X-ray emission has a nuclear origin and could be justified as the result of nuclear activity. This complements a set of low resolution observations (Oteo et al. 2017) in the mm bands for both SDP9 and SDP11 that suggest the presence of intense star formation could be associated with massive dense molecular gas reservoirs and high dense molecular gas fractions. 

On the basis of the X-ray data we attempted an estimate of star formation and BH properties for both our targets.  

Thus, by taking advantage of the lensing magnification we identified weak nuclear activity associated to high-$z$ galaxies with large star formation rates, useful to extend the investigation of the relationship between star formation and nuclear activity to two intrinsically less luminous, high-$z$ star forming galaxies than was possible so far. Given our results only for two objects, they cannot constrain the evolutionary models, but provide interesting hints and set an observational path towards addressing the role of star formation and nuclear activity in forming galaxies.

\begin{acknowledgements}
The scientific results reported in this article are based on observations made by the \chandra X-ray Observatory. This paper makes use of the following ALMA data: 2015.1.00415.S . ALMA is a partnership of ESO (representing its member states), NSF (USA) and NINS (Japan), together with NRC (Canada), NSC and ASIAA (Taiwan), and KASI (Republic of Korea), in cooperation with the Republic of Chile. The Joint ALMA Observatory is operated by ESO, AUI/NRAO and NAOJ. MM, MN, and AL acknowledge partial financial support by PRIN-INAF 2012 project ‘Looking into the dust-obscured phase of galaxy formation through cosmic zoom lenses in the Herschel Astrophysical Large Area Survey’.

MM and CM acknowledge partial financial support by the Italian {\it Ministero dell'Istruzione, Universit\`a e Ricerca} through the grant {\it Progetti Premiali 2012-iALMA} (CUP C52I13000140001).
AL, CM, and LD acknowledge partial support by PRIN INAF 2014 ‘Probing the AGN/galaxy co-evolution through ultra-deep and ultra-high-resolution radio surveys’. AL acknowledge partial support from the PRIN MIUR 2015 ‘Cosmology and Fundamental Physics: illuminating the Dark Universe with Euclid’ and from the RADIOFOREGROUNDS grant (COMPET-05-2015, agreement number 687312) of the European Union Horizon 2020 research and innovation programme. MN has received funding from the European Union's Horizon 2020 research and innovation programme under the Marie Sk{\l}odowska-Curie grant agreement No 707601. GDZ acknowledges partial financial support by ASI/INAF Agreement 2014-024-R.0 for the {\it Planck} LFI activity of Phase E2. 

We acknowledge J. Gonzales-Nuevo, M. Baes, A. Cooray, I. Oteo, D. A. Riechers, A. Omont, P. Ranalli, I. Georgantoupoulos, L. Fan, S. Serjeant, F. Carrera for the useful comments to some sections of the current paper and to the proposal. We acknowledge the work and coordination activity of the core team for the Herschel-ATLAS collaboration in the framework of which the targets were observed.
We acknowledge the referee, G. Rodighiero, for the useful comments. 
\end{acknowledgements}

%

\begin{thebibliography}{}
\bibitem[Alexander et al.(2008)]{2008AJ....135.1968A} Alexander, D.~M., Brandt, W.~N., Smail, I., et al.\ 2008, \aj, 135, 1968 
\bibitem[Aravena et al.(2016)]{2016MNRAS.457.4406A} Aravena, M., Spilker, J.~S., Bethermin, M., et al.\ 2016, \mnras, 457, 4406 
\bibitem[Arnaud(1996)]{1996ASPC..101...17A} Arnaud, K.~A.\ 1996, Astronomical Data Analysis Software and Systems V, 101, 17 
\bibitem[B{\'e}thermin et al.(2016)]{2016A&A...586L...7B} B{\'e}thermin, M., De Breuck, C., Gullberg, B., et al.\ 2016, \aap, 586, L7 
\bibitem[Bonzini et al.(2015)]{2015MNRAS.453.1079B} Bonzini, M., Mainieri, V., Padovani, P., et al.\ 2015, \mnras, 453, 1079 
\bibitem[Bussmann et al.(2013)]{2013ApJ...779...25B} Bussmann, R.~S., P{\'e}rez-Fournon, I., Amber, S., et al.\ 2013, \apj, 779, 25 
\bibitem[Cash(1979)]{1979ApJ...228..939C} Cash, W.\ 1979, \apj, 228, 939 
\bibitem[Citro et al.(2016)]{2016A&A...592A..19C} Citro, A., Pozzetti, L., Moresco, M., \& Cimatti, A.\ 2016, \aap, 592, A19 
\bibitem[Dye et al.(2014)]{2014MNRAS.440.2013D} Dye, S., Negrello, M., Hopwood, R., et al.\ 2014, \mnras, 440, 2013
\bibitem[Dye et al.(2015)]{2015MNRAS.452.2258D} Dye, S., Furlanetto, C., Swinbank, A.~M., et al.\ 2015, \mnras, 452, 2258 
\bibitem[Dannerbauer et al.(2014)]{2014A&A...570A..55D} Dannerbauer, H., Kurk, J.~D., De Breuck, C., et al.\ 2014, \aap, 570, A55 
\bibitem[Delvecchio et al.(2015)]{2015MNRAS.449..373D} Delvecchio, I., Lutz, D., Berta, S., et al.\ 2015, \mnras, 449, 373 
\bibitem[Eales et al.(2010)]{2010PASP..122..499E} Eales, S., Dunne, L., Clements, D., et al.\ 2010, \pasp, 122, 499 
\bibitem[Feroz et al.(2009)]{2009CQGra..26u5003F} Feroz, F., Gair, J.~R., Hobson, M.~P., \& Porter, E.~K.\ 2009, Classical and Quantum Gravity, 26, 215003 
\bibitem[Fritz et al.(2006)]{2006MNRAS.366..767F} Fritz, J., Franceschini, A., \& Hatziminaoglou, E.\ 2006, \mnras, 366, 767 
\bibitem[Fu et al.(2012)]{2012ApJ...753..134F} Fu, H., Jullo, E., Cooray, A., et al.\ 2012, \apj, 753, 134 
\bibitem[Hopkins et al.(2007)]{2007ApJ...662..110H} Hopkins, P.~F., Lidz, A., Hernquist, L., et al.\ 2007, \apj, 662, 110 
\bibitem[Imanishi et al.(2016)]{2016AJ....152..218I} Imanishi, M., Nakanishi, K., \& Izumi, T.\ 2016, \aj, 152, 218 
\bibitem[Johnson et al.(2013)]{2013MNRAS.431..662J} Johnson, S.~P., Wilson, G.~W., Wang, Q.~D., et al.\ 2013, \mnras, 431, 662 
\bibitem[Kawakatu et al.(2007)]{2007ApJ...663..924K} Kawakatu, N., Andreani, P., Granato, G.~L., \& Danese, L.\ 2007, \apj, 663, 924 
\bibitem[Kormann et al.(1994)]{1994A&A...284..285K} Kormann, R., Schneider, P., \& Bartelmann, M.\ 1994, \aap, 284, 285 
\bibitem[Lapi et al.(2014)]{2014ApJ...782...69L} Lapi, A., Raimundo, S., Aversa, R., et al.\ 2014, \apj, 782, 69 
\bibitem[Lapi et al.(2011)]{2011ApJ...742...24L} Lapi, A., Gonz{\'a}lez-Nuevo, J., Fan, L., et al.\ 2011, \apj, 742, 24 
\bibitem[Lilly et al.(2013)]{2013ApJ...772..119L} Lilly, S.~J., Carollo, C.~M., Pipino, A., Renzini, A., \& Peng, Y.\ 2013, \apj, 772, 119
\bibitem[Lupu et al.(2012)]{2012ApJ...757..135L} Lupu, R.~E., Scott, K.~S., Aguirre, J.~E., et al.\ 2012, \apj, 757, 135 
\bibitem[Maiolino(2008)]{2008NewAR..52..339M} Maiolino, R.\ 2008, \nar, 52, 339 
\bibitem[Mancuso et al.(2016)]{2016ApJ...833..152M} Mancuso, C., Lapi, A., Shi, J., et al.\ 2016, \apj, 833, 152 
\bibitem[Mullaney et al.(2012)]{2012ApJ...753L..30M} Mullaney, J.~R., Daddi, E., B{\'e}thermin, M., et al.\ 2012, \apjl, 753, L30 
\bibitem[Nayyeri et al.(2016)]{2016ApJ...823...17N} Nayyeri, H., Keele, M., Cooray, A., et al.\ 2016, \apj, 823, 17 
\bibitem[Nightingale \& Dye(2015)]{2015MNRAS.452.2940N} Nightingale, J.~W., \& Dye, S.\ 2015, \mnras, 452, 2940 
\bibitem[Negrello et al.(2017)]{2017MNRAS.465.3558N} Negrello, M., Amber, S., Amvrosiadis, A., et al.\ 2017, \mnras, 465, 3558 
\bibitem[Negrello et al.(2014)]{2014MNRAS.440.1999N} Negrello, M., Hopwood, R., Dye, S., et al.\ 2014, \mnras, 440, 1999 
\bibitem[Negrello et al.(2010)]{2010Sci...330..800N} Negrello, M., Hopwood, R., De Zotti, G., et al.\ 2010, Science, 330, 800 
\bibitem[Negrello et al.(2007)]{2007MNRAS.377.1557N} Negrello, M., Perrotta, F., Gonz{\'a}lez-Nuevo, J., et al.\ 2007, \mnras, 377, 1557 
\bibitem[Papovich et al.(2011)]{2011MNRAS.412.1123P} Papovich, C., Finkelstein, S.~L., Ferguson, H.~C., Lotz, J.~M., \& Giavalisco, M.\ 2011, \mnras, 412, 1123 
\bibitem[Piconcelli et al.(2005)]{2005A&A...432...15P} Piconcelli, E., Jimenez-Bail{\'o}n, E., Guainazzi, M., et al.\ 2005, \aap, 432, 15 
\bibitem[Popping et al.(2017)]{2017A&A...602A..11P} Popping, G., Decarli, R., Man, A.~W.~S., et al.\ 2017, \aap, 602, A11 
\bibitem[Ranalli et al.(2015)]{2015A&A...577A.121R} Ranalli, P., Georgantopoulos, I., Corral, A., et al.\ 2015, \aap, 577, A121 
\bibitem[Ranalli et al.(2003)]{2003A&A...399...39R} Ranalli, P., Comastri, A., \& Setti, G.\ 2003, \aap, 399, 39 
\bibitem[Rodighiero et al.(2015)]{2015ApJ...800L..10R} Rodighiero, G., Brusa, M., Daddi, E., et al.\ 2015, \apjl, 800, L10 
\bibitem[Rodighiero et al.(2014)]{2014MNRAS.443...19R} Rodighiero, G., Renzini, A., Daddi, E., et al.\ 2014, \mnras, 443, 19 
\bibitem[Rodighiero et al.(2011)]{2011ApJ...739L..40R} Rodighiero, G., Daddi, E., Baronchelli, I., et al.\ 2011, \apjl, 739, L40 
\bibitem[Smit et al.(2012)]{2012ApJ...756...14S} Smit, R., Bouwens, R.~J., Franx, M., et al.\ 2012, \apj, 756, 14 
\bibitem[Spaans \& Meijerink(2008)]{2008ApJ...678L...5S} Spaans, M., \& Meijerink, R.\ 2008, \apjl, 678, L5 
\bibitem[Speagle et al.(2014)]{2014ApJS..214...15S} Speagle, J.~S., Steinhardt, C.~L., Capak, P.~L., \& Silverman, J.~D.\ 2014, \apjs, 214, 15 
\bibitem[Suyu et al.(2006)]{2006MNRAS.371..983S} Suyu, S.~H., Marshall, P.~J., Hobson, M.~P., \& Blandford, R.~D.\ 2006, \mnras, 371, 983 
\bibitem[Wang et al.(2013)]{2013ApJ...773...44W} Wang, R., Wagg, J., Carilli, C.~L., et al.\ 2013, \apj, 773, 44 
\bibitem[Wardlow et al.(2013)]{2013ApJ...762...59W} Wardlow, J.~L., Cooray, A., De Bernardis, F., et al.\ 2013, \apj, 762, 59 
\bibitem[Warren \& Dye(2003)]{2003ApJ...590..673W} Warren, S.~J., \& Dye, S.\ 2003, \apj, 590, 673 
\bibitem[Wong et al.(2017)]{2017ApJ...843L..35W} Wong, K.~C., Ishida, T., Tamura, Y., et al.\ 2017, \apjl, 843, L35 
\bibitem[Wright et al.(2010)]{2010AJ....140.1868W} Wright, E.~L., Eisenhardt, P.~R.~M., Mainzer, A.~K., et al.\ 2010, \aj, 140, 1868-1881 
\end{thebibliography}
%

\end{document}